\documentclass[11pt]{article}

\usepackage{times}
\usepackage{float}

\usepackage{amsfonts,amstext,amsmath,amssymb,amsthm}

\usepackage{chngpage}
\usepackage{rotating}
\usepackage{graphicx}
\usepackage{epstopdf} % to be loaded after the 'graphicx' package

\usepackage{caption}
\usepackage{subcaption}
\usepackage{mathrsfs}

\long\def\symbolfootnote[#1]#2{\begingroup%
\def\thefootnote{\fnsymbol{footnote}}\footnote[#1]{#2}\endgroup}
\usepackage{hyperxmp}
\usepackage[%
    bookmarks=true,         % show bookmarks bar?
    unicode=false,          % non-Latin characters in Acrobat’s bookmarks
    pdftoolbar=true,        % show Acrobat’s toolbar?
    pdfmenubar=true,        % show Acrobat’s menu?
    hyperfootnotes=false,   % suppress hyperref's hyperfootnotes
    pdffitwindow=false,     % window fit to page when opened
    pdfstartview={FitH},    % fits the width of the page to the window
    pdftitle={On the Rate of Convergence of the Quasi- to Stationary Distribution for the Shiryaev--Roberts Diffusion},    % title
    pdfauthor={Kexuan Li,Aleksey S. Polunchenko},     % author
    pdfsubject={On the Rate of Convergence of the Quasi- to Stationary Distribution for the Shiryaev--Roberts Diffusion},   % subject of the document
    pdfcreator={Aleksey S. Polunchenko},   % creator of the document
    pdfkeywords={First exit times,Generalized Shiryaev--Roberts procedure,Laplace transform,Markov diffusions,Moment generating function,Quickest change-point detection,Whittaker functions}, % list of keywords
    pdfnewwindow=true,      % links in new window
    colorlinks=false,       % false: boxed links; true: colored links
    linkcolor=red,          % color of internal links (change box color with linkbordercolor)
    citecolor=green,        % color of links to bibliography
    filecolor=magenta,      % color of file links
    urlcolor=cyan           % color of external links
]{hyperref}

\usepackage{suffix}
\usepackage{mathtools}

\usepackage{paralist}

\usepackage[T1]{fontenc}

\usepackage{titlesec} % Very fancy section title control: Sets sections as SQA requires

\titleformat{\section}{\large\bfseries\uppercase}{\thesection.}{.5em}{}
\titlespacing*{\section}{0pt}{*3}{*2}
\titleformat{\subsection}{\normalfont\bfseries}{\thesubsection.}{.5em}{}
\titlespacing*{\subsection}{0pt}{*3}{*2}
\titleformat{\subsubsection}{\normalfont\bfseries}{\thesubsubsection.}{.5em}{}
\titlespacing*{\subsubsection}{0pt}{*3}{*2}

\numberwithin{equation}{section}

\usepackage[sort&compress]{natbib}

\usepackage{bigints}

\DeclarePairedDelimiterX\MeijerM[3]{\lparen}{\rparen}%
{\begin{smallmatrix}#1 \\ #2\end{smallmatrix}\delimsize\vert\,#3}

\newcommand\MeijerG[8][]{%
  G^{\,#2,#3}_{#4,#5}\MeijerM[#1]{#6}{#7}{#8}}

\WithSuffix\newcommand\MeijerG*[7]{%
  G^{\,#1,#2}_{#3,#4}\MeijerM*{#5}{#6}{#7}}

\long\def\symbolfootnote[#1]#2{\begingroup%
\def\thefootnote{\fnsymbol{footnote}}\footnote[#1]{#2}\endgroup}

\renewcommand{\Pr}{\mathbb{P}} % probability
\DeclareMathOperator{\EV}{\mathbb{E}} % expected value

\DeclareMathOperator{\Ei}{Ei}
\DeclareMathOperator{\E1}{E_1}

\renewcommand{\le}{\leqslant} % AMS le ge
\renewcommand{\ge}{\geqslant}

\newcommand{\abs}[1]{\left\vert#1\right\vert}

\DeclareMathOperator{\One}{\mathchoice{\rm 1\mskip-4.2mu l}{\rm 1\mskip-4.2mu l}{\rm 1\mskip-4.6mu l}{\rm 1\mskip-5.2mu l}}
\newcommand{\indicator}[1]{{\One_{\left\{#1\right\}}}}

\theoremstyle{plain} %% produces italic text

\newtheorem{remark}{Remark}[section]

\graphicspath{{./gfx/}}
\usepackage[norefs,nocites]{refcheck}
\usepackage{silence}
\begin{document}

\title{\textbf{\Large On the Convergence Rate of the Quasi- to Stationary Distribution\\ for the Shiryaev--Roberts Diffusion}}

\date{}
\maketitle

%%%%%%%%% Authors, affiliations %%%%%%%%%%%%%%%%%%%%%%%%%%
\begin{center}
\null\vskip-2cm\author{
\textbf{\large Kexuan Li\ and\ Aleksey\ S.\ Polunchenko}\\
Department of Mathematical Sciences, State University of New York at Binghamton,\\Binghamton, New York, USA
}
\end{center}
%
%-------------------------------------------------------------------------------------------------%
%
% At the bottom of the first page, provide the e-mail, telephone and fax
% info, and complete address (including the street name, P. O. Box number,
% etc. as needed) for the corresponding author, formatted as follows:
%
% Address correspondence to D. H. Author, Department of Statistics, CLAS
% Building U-4120, University of Connecticut, 215 Glenbrook Road, Storrs, CT % 06269-4120, USA; Fax: 860-486-4113; E-mail: nitis.mukhopadhyay@uconn.edu
%
\symbolfootnote[0]{\normalsize\hspace{-0.6cm}Address correspondence to A.\ S.\ Polunchenko, Department of Mathematical Sciences, State University of New York (SUNY) at Binghamton, 4400 Vestal Parkway East, Binghamton, NY 13902--6000, USA; Tel: +1 (607) 777--6906; Fax: +1 (607) 777--2450; E-mail:~\href{mailto:aleksey@binghamton.edu}{aleksey@binghamton.edu}.}\\
%
%-------------------------------------------------------------------------------------------------%
%
{\small\noindent\textbf{Abstract:} For the classical Shiryaev--Roberts martingale diffusion considered on the interval $[0,A]$, where $A>0$ is a given absorbing boundary, it is shown that the rate of convergence of the diffusion's quasi-stationary cumulative distribution function (cdf), $Q_{A}(x)$, to its stationary cdf, $H(x)$, as $A\to+\infty$, is no worse than $O(\log(A)/A)$, uniformly in $x\ge0$. The result is established explicitly, by constructing new tight lower- and upper-bounds for $Q_{A}(x)$ using certain latest monotonicity properties of the modified Bessel $K$ function involved in the exact closed-form formula for $Q_{A}(x)$ recently obtained by~\cite{Polunchenko:SA2017a}.
 % New lower- and upper-bounds for $Q_{A}(x)$ are offered as well. % As an application to quickest change-point detection, the randomized Shiryaev--Roberts--Pollak procedure known to be nearly Pollak-minimax in the limit as $A\to+\infty$, is shown to approach the unknown optimum no slower than $O(??)$. This is an improvement over the earlier result of~\cite{Polunchenko:TPA2017} who showed the speed of convergence to be at least $O(1/\sqrt{A})$.
}
\\ \\
%-------------------------------------------------------------------------------------------------%
% KEYWORDS
%
% Keywords are to be listed alphabetically (with the first letter
% capitalized), preferably chosen from the text and not from the title of the % paper itself; the importance of the words used in a title is already
% obvious to readers. The keywords are to be separated by semicolons (;).
%
{\small\noindent\textbf{Keywords:} {Generalized Shiryaev--Roberts procedure; Markov diffusion; Quickest change-point detection; Whittaker functions.}
\\ \\
%-------------------------------------------------------------------------------------------------%
% SUBJECT CLASSIFICATIONS
%
% MSC2010, see http://www.ams.org/msc/
%
%  Statistics -> Sequential methods
%   62L10 - Sequential analysis
%   62L15 - Optimal stopping
%
%  Probability theory and stochastic processes -> Stochastic processes
%   60G10 - Stationary processes
%   60G40 - Stopping times; optimal stopping problems; gambling theory
%
%  Statistics -> Parametric inference
%   62F12 - Asymptotic properties of estimators
%   62F15 - Bayesian inference
%
%  Statistics -> Decision theory
%   62C10 - Bayesian problems; characterization of Bayes procedures
%   62C20 - Minimax procedures
%
%  Statistics -> Inference from stochastic processes
%   62M15 - Spectral analysis
%
%  Probability theory and stochastic processes -> Markov processes
% 	60J25 - Continuous-time Markov processes on general state spaces
%   60J60 - Diffusion processes
%   60J65 - Brownian motion
%
%  Special functions
%   33C15 - Confluent hypergeometric functions, Whittaker functions, ${}_1F_1$
%

%+-----------------------------------------------------------------------------------------------+%
\section{Introduction} % Initial capital letter, then lower case. No full stop.
\label{sec:intro}

This work is an attempt to quantify the relationship between the phenomena of quasi-stationarity and stationarity exhibited by one particular version of the Generalized Shiryaev--Roberts (GSR) stochastic process---a time-homogeneous Markov diffusion well-known in the area of quickest change-point detection. See, e.g.,~\cite{Shiryaev:SMD61,Shiryaev:TPA63,Shiryaev:Book78,Shiryaev:Bachelier2002,Shiryaev:Book2011,Shiryaev:Book2017},~\cite{Pollak+Siegmund:B85},~\cite{Feinberg+Shiryaev:SD2006},~\cite{Burnaev+etal:TPA2009},~\cite{Polunchenko+Sokolov:MCAP2016}, and~\cite{Polunchenko:SA2016,Polunchenko:SA2017a,Polunchenko:SA2017b,Polunchenko:TPA2017}. More specifically, the GSR process' version of interest is the solution $(R_{t}^{r})_{t\ge0}$ of the stochastic differential equation (SDE)
\begin{align}\label{eq:Rt_r-def}
dR_{t}^{r}
&=
dt+R_{t}^{r} dB_{t}
\;\text{with}\;
R_{0}^{r}\coloneqq r\ge0,
\end{align}
where $(B_{t})_{t\ge0}$ is standard Brownian motion in the sense that $\EV[dB_t]=0$, $\EV[(dB_t)^2]=dt$, and $B_0=0$; the initial value $R_{0}^{r}\coloneqq r$ is often referred to as the process' headstart. It is straightforward to solve~\eqref{eq:Rt_r-def} and express $(R_{t}^{r})_{t\ge0}$ explicitly as
\begin{align}
R_{t}^{r}
&=
\exp\left\{B_{t}-\dfrac{1}{2}t\right\}\left(r+\bigintsss_{0}^{t} \exp\biggl\{-\left(B_s-\dfrac{1}{2}s\right)\biggr\}ds\right),
\;\;
t\ge0,
\end{align}
so that the set $[0,+\infty)$ is easily seen to be the ``natural'' state space for $(R_{t}^{r})_{t\ge0}$ because $R_0^{r}\coloneqq r\ge0$. Moreover, it is also easily checked that $\EV[R_{t}^{r}-t-r]=0$ for any $t,r\ge0$, i.e., the process $\{R_{t}^{r}-t-r\}_{t\ge0}$ is a zero-mean martingale. Yet, although $(R_{t}^{r})_{t\ge0}$ has a linear upward trend in time, it is actually a recurrent process with a nontrivial probabilistic behavior in the limit, as $t\to+\infty$; cf.~\cite[p.~270]{Pollak+Siegmund:B85}. Specifically, if $(R_{t}^{r})_{t\ge0}$ is let run ``loose'', i.e., considered on the entire nonnegative half-line, then the limiting (as $t\to+\infty$) behavior of $(R_{t}^{r})_{t\ge0}$ is known as stationarity. The latter is characterized by the invariant probability measure whose cumulative distribution function (cdf) and density (pdf), respectively, are
\begin{align}\label{eq:SR-StDist-def}
H(x)
&\coloneqq
\lim_{t\to+\infty}\Pr(R_{t}^{r}\le x)
\;\;
\text{and}
\;\;
h(x)
\coloneqq
\dfrac{d}{dx}H(x),
\end{align}
provided $r\in[0,+\infty)$. This probability measure has already been found, e.g., by~\cite{Shiryaev:SMD61,Shiryaev:TPA63}, by~\cite{Pollak+Siegmund:B85}, and more recently also by~\cite{Feinberg+Shiryaev:SD2006,Burnaev+etal:TPA2009,Polunchenko+Sokolov:MCAP2016}, to be the momentless (no moments of orders one and higher) distribution
\begin{align}\label{eq:SR-StDist-answer}
H(x)
&=
e^{-\tfrac{2}{x}}\indicator{x\ge0}
\;\;
\text{and}
\;\;
h(x)
=
\dfrac{2}{x^2}\,e^{-\tfrac{2}{x}}\indicator{x\ge0},
\end{align}
which is an extreme-value Fr\'{e}chet-type distribution, and a particular case of the inverse (reciprocal) gamma distribution. See also, e.g.,~\cite{Linetsky:OR2004} and~\cite{Avram+etal:MPRF2013}. As an aside, note that, in view of~\eqref{eq:SR-StDist-answer}, the stationary distribution of the reciprocal of $(R_{t}^{r})_{t\ge0}$ is exponential with mean $1/2$.

However, if all states from a fixed $A>0$ and up inside the process' ``natural'' state space $[0,+\infty)$ are made into absorbing states, then $(R_{t}^{r})_{t\ge0}$ also has a nontrivial probabilistic behavior in the limit as $t\to+\infty$. This behavior is known as quasi-stationarity, and it is characterized by the invariant probability measure whose cdf and pdf, respectively, are
\begin{align}\label{eq:QSD-def}
Q_{A}(x)
&
\coloneqq
\lim_{t\to+\infty}\Pr(R_{t}^{r}\le x|R_{s}^{r}\in[0,A)\;\text{for all}\;0\le s\le t)
\;\;
\text{and}
\;\;
q_{A}(x)
\coloneqq
\dfrac{d}{dx}Q_{A}(x),
\end{align}
provided $r\in[0,A)$. The existence of this probability measure was formally established, e.g., by~\cite{Pollak+Siegmund:B85}, although one can also infer the same result, e.g., from the earlier seminal work of~\cite{Mandl:CMJ1961}. Moreover, analytic closed-form formulae for $Q_{A}(x)$ and $q_{A}(x)$ were recently obtained by~\cite{Polunchenko:SA2017a}, apparently for the first time in the literature; see formulae~\eqref{eq:QSD-pdf-answer} and~\eqref{eq:QSD-cdf-answer-W0} in Section~\ref{sec:main-results} below. These formulae were used by~\cite{Polunchenko+Pepelyshev:SP2018} to compute analytically the quasi-stationary distribution's Laplace transform, and then also by~\cite{Li+etal:CommStat2019} to find the quasi-stationary distribution's fractional moment of any real order.
\begin{remark}
The phenomenon of quasi-stationarity is also exhibited by $(R_{t}^{r})_{t\ge0}$ in another case, viz. when all states from 0 up through a fixed $A>0$ inclusive inside the process' ``natural'' state space $[0,+\infty)$ are made into absorbing states, so that the state space of $(R_{t}^{r})_{t\ge0}$ becomes the set $[A,+\infty)$ with absorbtion at the lower end. This case was first investigated in~\cite[Section~7.8.2]{Collet+etal:Book2013}. It was also recently analyzed by~\cite{Polunchenko+etal:TPA2018} who obtained analytically exact closed-form formulae for the quasi-stationary cdf and pdf.
\end{remark}

Not surprisingly, the quasi-stationary distribution~\eqref{eq:QSD-def} and the stationary distribution~\eqref{eq:SR-StDist-def} {\em are} related: as one would expect, the former converges to the latter as $A\to+\infty$. This was formally shown by~\cite{Pollak+Siegmund:JAP1996}, and not only for the GSR process at hand, but for an entire class of stochastically monotone processes. More specifically, it can be deduced from~\cite{Pollak+Siegmund:JAP1996} that $Q_{A}(x)\ge H(x)$ for any fixed $A>0$ and any $x\ge 0$, and that $\lim_{A\to+\infty} Q_{A}(x)=H(x)$ for any fixed $x\ge0$. The principal question addressed in this work is that of the actual rate of convergence of $Q_{A}(x)$ down to $H(x)$ as $A\to+\infty$ uniformly in $x\ge0$: the answer---obtained and reported in Section~\ref{sec:main-results}---is $(0<)\;\sup_{x\ge0}[Q_{A}(x)-H(x)]=O(\log(A)/A)$. This is the main contribution of this work, but not its only contribution: our $Q_{A}(x)$-to-$H(x)$ convergence (as $A\to+\infty$) analysis relies on new lower- and upper-bounds for $Q_{A}(x)$; the upperbounds are three, of varying complexity and accuracy, and all are much tighter than the trivial $Q_{A}(x)\le 1$, while the lowerbound is one and it is much tighter than $H(x)\le Q_{A}(x)$ implied by the earlier work of~\cite{Pollak+Siegmund:JAP1996}. All of the bounds, which may be considered another contribution of this work, are obtained explicitly with the aid of the formula for $Q_{A}(x)$ latterly offered by~\cite{Polunchenko:SA2017a}, and certain recently discovered monotonicity properties of the modified Bessel $K$ function (of the second kind); the latter is involved in the formula for $Q_{A}(x)$ obtained by~\cite{Polunchenko:SA2017a}.

The process $(R_{t}^{r})_{t\ge0}$ governed by equation~\eqref{eq:Rt_r-def} arises in quickest change-point detection when the aim is to control the mean of the process $X_{t}\coloneqq t\indicator{t>\nu}+B_{t}$, where $\nu,t\ge0$, observed ``live''. Since $\EV[X_{t}]=t\indicator{t>\nu}$, it is anticipated that the drift of $(X_{t})_{t\ge0}$ will change from none (zero) to one (per time unit) at time instance $\nu\in[0,+\infty]$ referred to as the change-point. The challenge is that $\nu$ is not known in advance; in particular $\nu=\infty$ is a possibility, i.e., the drift of $(X_{t})_{t\ge0}$ may remain zero indefinitely and never change. The mean of $(X_{t})_{t\ge0}$ is controlled by sounding an alarm should and as soon as the behavior of $(X_{t})_{t\ge0}$ suggest that possibly $\EV[X_{t}]= t\neq 0$, i.e., $t>\nu$; if it is not the case, then the alarm is a false one. More concretely, the so-called GSR quickest change-point detection procedure, set up to control the drift of $(X_t)_{t\ge0}$, sounds a false alarm at
\begin{align}\label{eq:T-GSR-def}
\mathcal{S}_{A}^{r}
&\coloneqq
\inf\big\{t\ge0\colon R_{t}^{r}=A\big\},
\;
r\in[0,A),
\end{align}
where the constant $A>0$ is selected in advance in accordance with the desired false alarm risk level; it is to be understood in the right-hand side of the definition of $\mathcal{S}_{A}^{r}$ that $\inf\{\varnothing\}=+\infty$. Hence $(R_{t}^{r})_{t\ge0}$ is the GSR procedure's detection statistic in the pre-change regime, i.e., for $t\in[0,\nu]$. It is of note that $\Pr(\mathcal{S}_{A}^{r}<+\infty)=1$. The definition~\eqref{eq:QSD-def} of the quasi-stationary cdf can be rewritten as $Q_{A}(x)=\lim_{t\to+\infty}\Pr(R_{t}^{r}\le x|\mathcal{S}_{A}^{r}>t)$.

The aforementioned GSR procedure was proposed by~\cite{Moustakides+etal:SS11} as a headstarted (i.e., more general) version of the classical quasi-Bayesian Shiryaev--Roberts (SR) procedure that emerged from the independent work of Shiryaev~\citeyearpar{Shiryaev:SMD61,Shiryaev:TPA63} and that of Roberts~\citeyearpar{Roberts:T66}. The interest in the GSR procedure (and its variations) is due to its strong (near-) optimality properties. See, e.g.,~\cite{Burnaev:ARSAIM2009},~\cite{Feinberg+Shiryaev:SD2006},~\cite{Burnaev+etal:TPA2009},~\cite{Polunchenko+Tartakovsky:AS10},~\cite{Tartakovsky+Polunchenko:IWAP10},~\cite{Vexler+Gurevich:MESA2011}, and~\cite{Tartakovsky+etal:TPA2012}. For example, it is known that if the GSR procedure's headstart is sampled from the quasi-stationary distribution~\eqref{eq:QSD-def}, then such a randomization of the GSR procedure makes the latter nearly (to within a vanishingly small additive term) minimax in the sense of~\cite{Pollak:AS85}. The idea of such a randomization of the GSR procedure and a proof that the randomized GSR procedure is nearly minimax are due to~\cite{Pollak:AS85} who was concerned with the discrete-time formulation of the problem. For the problem's continuous-time formulation, the same result was established by~\cite{Polunchenko:TPA2017} who heavily relied on the exact closed-form formulae for $Q_{A}(x)$ and $q_{A}(x)$ obtained by~\cite{Polunchenko:SA2016}, as well as on the quasi-stationary distribution's first two moments, also computed by~\cite{Polunchenko:SA2016}.

The rest of the paper is three sections. The first one, Section~\ref{sec:nomenclature}, is to introduce our notation and to provide the necessary preliminary background on the special functions needed for our convergence analysis. The next section, Section~\ref{sec:main-results}, is the paper's main section: this is where we formally state and prove our main result. Lastly, in Section~\ref{sec:conclusion} we make a few concluding remarks and wrap up the entire paper.

%+-----------------------------------------------------------------------------------------------+%
\section{Notation and nomenclature}
\label{sec:nomenclature}

For ease of exposition, we shall follow the standard notation employed uniformly across mathematical literature. This includes not only the usual symbols $\mathbb{R}$, $\mathbb{C}$, $\mathbb{N}$, $\mathbb{Z}$, $\mathrm{i}\coloneqq\sqrt{-1}$, and so on, but, more importantly, also an array of special functions we are to deal with throughout the sequel. These functions, in their most common notation, are:
\begin{enumerate}
    \setlength{\itemsep}{10pt}
    \setlength{\parskip}{0pt}
    \setlength{\parsep}{0pt}
    \item The Gamma function $\Gamma(z)$, where $z\in\mathbb{C}$, sometimes also regarded as the extension of the factorial to complex numbers, due to the property $\Gamma(n)=(n-1)!$ exhibited for $n\in\mathbb{N}$. See, e.g.,~\cite[Chapter~1]{Bateman+Erdelyi:Book1953v1}.
    \item The exponential integral function $\Ei(x)$, where $x\in\mathbb{R}\backslash\{0\}$, defined as
    \begin{align}\label{eq:Ei-func-def}
    \Ei(x)
    &\coloneqq%
    \begin{cases}
    -\displaystyle\int_{-x}^\infty\dfrac{e^{-y}}{y}\,dy,&\text{if $x<0$;}\\[4mm]
    -\lim_{\varepsilon\to+0}\left[\displaystyle\int_{-x}^{-\varepsilon}\dfrac{e^{-y}}{y}\,dy+\displaystyle\int_{\varepsilon}^{\infty}\dfrac{e^{-y}}{y}\,dy\right],&\text{if $x>0$},
    \end{cases}
    \end{align}
    with a singularity at $x=0$. Its basic properties are summarized, e.g., in~\cite[Chapter~5]{Abramowitz+Stegun:Handbook1964}. More specifically, we will need the function $\E1(x)\coloneqq-\Ei(-x)$ with $x>0$, so that
    \begin{align}\label{eq:E1-func-def}
    \E1(x)
    &\coloneqq
    \int_{x}^\infty\dfrac{e^{-y}}{y}\,dy
    =
    x e^{-x}\int_{0}^{+\infty} e^{-xy}\log(1+y)\,dy,\; x>0,
    \end{align}
    where the second equality is because
    \begin{align*}
    \int_{0}^{+\infty} e^{-ay}\log(b+y)\,dy
    &=
    \dfrac{1}{a}\big[\log b-e^{ab}\Ei(-ba)\big],
    \;
    \abs{\,\arg(b)}<\pi,\; \Re(a)>0,
    \end{align*}
    as given, e.g., by~\cite[Integral~4.337.1,~p.~572]{Gradshteyn+Ryzhik:Book2007}.
    \item The Whittaker $M$ and $W$ functions, traditionally denoted, respectively, as $M_{a,b}(z)$ and $W_{a,b}(z)$, where $a,b,z\in\mathbb{C}$; the Whittaker $M$ function is undefined when $-2b\in\mathbb{N}$, but can be regularized. These functions were introduced by Whittaker~\citeyearpar{Whittaker:BAMS1904} as the fundamental solutions to the Whittaker differential equation. See, e.g.,~\cite{Slater:Book1960} and~\cite{Buchholz:Book1969}.
    \item The modified Bessel functions of the first and second kinds, conventionally denoted, respectively, as $I_{a}(z)$ and $K_{a}(z)$, where $a,z\in\mathbb{C}$; the index $a$ is referred to as the function's order. See~\cite[Chapter~7]{Bateman+Erdelyi:Book1953v2}. These functions form a set of fundamental solutions to the modified Bessel differential equation. The modified Bessel $K$ function is also known as the MacDonald function.
\end{enumerate}

%%%%%%%%%%%%%%%%%%%%%%

%+-----------------------------------------------------------------------------------------------+%
\section{Analysis of the rate of convergence}
\label{sec:main-results}

To set the ground for our quasi- to stationary distribution convergence analysis we begin by recalling a few earlier results due to~\cite{Polunchenko:SA2017a}. Specifically, it can be deduced from~\cite[Theorem~3.1]{Polunchenko:SA2017a} that if $A>0$ is fixed and $\lambda\equiv\lambda_A>0$ is the smallest (positive) solution of the equation
\begin{align}\label{eq:lambda-eqn}
W_{1,\tfrac{1}{2}\xi(\lambda)}\left(\dfrac{2}{A}\right)
&=
0,
\end{align}
where
\begin{align}\label{eq:xi-def}
\xi(\lambda)
&\coloneqq
\sqrt{1-8\lambda}
\;\;
\text{so that}
\;\;
\lambda
=
\dfrac{1}{8}\left(1-\big[\xi(\lambda)\big]^2\right),
\end{align}
then the quasi-stationary pdf is given by
\begin{align}\label{eq:QSD-pdf-answer}
q_A(x)
&=
\dfrac{e^{-\tfrac{1}{x}}\,\dfrac{1}{x}\,W_{1,\tfrac{1}{2}\xi(\lambda)}\left(\dfrac{2}{x}\right)}{e^{-\tfrac{1}{A}}\,W_{0,\tfrac{1}{2}\xi(\lambda)}\left(\dfrac{2}{A}\right)}\indicator{x\in[0,A]},
\end{align}
and the respective cdf is given by
\begin{align}\label{eq:QSD-cdf-answer-W0}
Q_A(x)
&=
\begin{cases}
1,&\;\text{if $x\ge A$;}\\[2mm]
\dfrac{e^{-\tfrac{1}{x}}\,W_{0,\tfrac{1}{2}\xi(\lambda)}\left(\dfrac{2}{x}\right)}{e^{-\tfrac{1}{A}}\,W_{0,\tfrac{1}{2}\xi(\lambda)}\left(\dfrac{2}{A}\right)},&\;\text{if $x\in[0,A)$;}\\[8mm]
0,&\;\text{otherwise},
\end{cases}
\end{align}
and $q_A(x)$ and $Q_A(x)$ are each a sufficiently smooth function of $x\ge0$ and $A>0$; observe from~\eqref{eq:lambda-eqn}, \eqref{eq:xi-def}, and~\eqref{eq:QSD-pdf-answer} that $q_A(A)=0$. The smoothness of $q_A(x)$ and $Q_A(x)$ is due to certain analytic properties of the Whittaker $W$ function on the right of formulae~\eqref{eq:QSD-pdf-answer} and~\eqref{eq:QSD-cdf-answer-W0}. These formulae stem from the solution of a certain Sturm--Liouville problem, and $\lambda$ is the smallest positive eigenvalue of the corresponding Sturm--Liouville operator; if the Sturm--Liouville operator is negated, as was done by~\cite{Polunchenko:SA2017a}, then $\lambda$ becomes the operator's largest {\em negative} eigenvalue.
\begin{remark}\label{rem:xi-symmetry}
The definition~\eqref{eq:xi-def} of $\xi(\lambda)$ can actually be changed to $\xi(\lambda)\coloneqq -\sqrt{1-8\lambda}$ with no effect whatsoever on either equation~\eqref{eq:lambda-eqn}, or formulae~\eqref{eq:QSD-pdf-answer} and~\eqref{eq:QSD-cdf-answer-W0}, i.e., all three are invariant with respect to the sign of $\xi(\lambda)$. This was previously pointed out by~\cite{Polunchenko:SA2017a}, and the reason for this $\xi(\lambda)$-symmetry is because equation~\eqref{eq:lambda-eqn} and formulae~\eqref{eq:QSD-pdf-answer} and~\eqref{eq:QSD-cdf-answer-W0} each have $\xi(\lambda)$ present only as (double) the second index of the corresponding Whittaker $W$ function or functions involved, and the Whittaker $W$ function in general is known (see, e.g.,~\cite[Identity~(19),~p.~19]{Buchholz:Book1969}) to be an even function of its second index, i.e., $W_{a,b}(z)=W_{a,-b}(z)$.
\end{remark}

It is evident that equation~\eqref{eq:lambda-eqn} is a key component of formulae~\eqref{eq:QSD-pdf-answer} and~\eqref{eq:QSD-cdf-answer-W0}, and consequently, of all of the characteristics of the quasi-stationary distribution as well. As a transcendental equation, it can only be solved numerically, although to within any desired accuracy, as was previously done, e.g., by~\cite{Linetsky:OR2004,Polunchenko:SA2016,Polunchenko:SA2017a,Polunchenko:SA2017b}, with the aid of {\it Mathematica} developed by Wolfram Research: {\it Mathematica}'s
special functions capabilities have long proven to be superb. Yet, it is known (see, e.g.,~\citealt{Linetsky:OR2004} and~\citealt{Polunchenko:SA2016}) that for any fixed $A>0$, the equation has countably many simple solutions $0<\lambda_1<\lambda_2<\lambda_3<\cdots$, such that $\lim_{k\to+\infty}\lambda_k=+\infty$. All of them depend on $A$, but since we are interested only in the smallest one, we shall use either the ``short'' notation $\lambda$, or the more explicit $\lambda_A$ to emphasize the dependence on $A$. It was shown by~\cite{Polunchenko:SA2016} that $\lambda_A$ is a monotonically decreasing function of $A$, such that
\begin{align}\label{eq:lambda-dbl-ineq}
\dfrac{1}{A}
+
\dfrac{1}{A(1+A)}
&<
\lambda_{A}
<
\dfrac{1}{A}
+
\dfrac{1+\sqrt{4A+1}}{2A^2}
,
\;\;
\text{for any}
\;\;
A>0,
\end{align}
whence $\lim_{A\to+\infty}\lambda_A=0$, and more specifically $\lambda_A=A^{-1}+O(A^{-3/2})$; cf.~\cite[p.~136 and Lemma~3.3]{Polunchenko:SA2016}. See also~\cite{Polunchenko+Pepelyshev:SP2018} for a discussion of potential ways to improve the foregoing double inequality.
\begin{remark}\label{rem:xi-complex-real}
Since $\lambda\equiv\lambda_{A}$ is monotonically decreasing in $A$, and such that $\lim_{A\to+\infty}\lambda_{A}=0$, one can conclude from~\eqref{eq:xi-def} that $\xi(\lambda_{A})$, for any finite $A>0$, is either \begin{inparaenum}[\itshape(a)]\item purely imaginary (i.e., $\xi(\lambda)=\mathrm{i}\alpha$ where $\mathrm{i}\coloneqq\sqrt{-1}$ and $\alpha\in\mathbb{R}$) if $A$ is sufficiently small, or \item purely real and between 0 inclusive and 1 exclusive (i.e., $0\le \xi(\lambda)<1$) otherwise\end{inparaenum}. The borderline case is when $\xi(\lambda)=0$, i.e., when $\lambda_{A}=1/8$, and the corresponding critical value of $A$ is the solution $\tilde{A}>0$ of the equation
\begin{align}\label{eq:tildeA-eqn-def}
W_{1,0}\left(\dfrac{2}{\tilde{A}}\right)
&=
0,
\;\;
\text{so that}
\;\;
\tilde{A}\approx10.240465,
\end{align}
as can be established by a basic numerical calculation. Hence, if $A<\tilde{A}\approx10.240465$, then $\lambda_{A}>1/8$ so that $\xi(\lambda)$ is purely imaginary; otherwise, if $A\ge\tilde{A}\approx10.240465$, then $\lambda_{A}\in(0,1/8]$ so that $\xi(\lambda)$ is purely real and such that $\xi(\lambda)\in[0,1)$ with $\lim_{A\to+\infty}\xi(\lambda_{A})=1$.
\end{remark}

Since the SR process is a stochastically monotone Markov process, it can be concluded at once from~\cite{Pollak+Siegmund:JAP1996} that $Q_{A}(x)\downarrow H(x)$ as $A\to+\infty$ for every fixed $x\ge0$. However, since $Q_{A}(x)$ is given {\em explicitly} by formula~\eqref{eq:QSD-cdf-answer-W0}, it is of interest to see if the same conclusion can be reached in a more {\em explicit} fashion. To this end, recall first the limit
\begin{align*}
\lim_{A\to+\infty}\left\{e^{-\tfrac{1}{A}}\,W_{0,\tfrac{1}{2}\xi(\lambda_{A})}\left(\dfrac{2}{A}\right)\right\}
&=1,
\end{align*}
which was previously proved by~\cite{Polunchenko:SA2017b}. On account of this limit it can be seen from~\eqref{eq:QSD-cdf-answer-W0} that
\begin{align*}
Q_{\infty}(x)
&\coloneqq
\lim_{A\to+\infty}Q_{A}(x)
=
e^{-\tfrac{1}{x}}W_{0,\tfrac{1}{2}}\left(\dfrac{2}{x}\right),
\;\;
\text{for every}
\;\;
x\ge0,
\end{align*}
because $\xi(\lambda_{A})\to 1$, as $A\to+\infty$, by Remark~\ref{rem:xi-complex-real}. Now, since
\begin{align}\label{eq:Whit-func-exp-id}
W_{a,a-\tfrac{1}{2}}(z)
&=
z^{a}e^{-\tfrac{z}{2}}
=
W_{a,\tfrac{1}{2}-a}(z),
\end{align}
as given, e.g., by~\cite[Identity~(28a),~p.~23]{Buchholz:Book1969}, so that
\begin{align*}
W_{0,\tfrac{1}{2}}(z)
&=
e^{-\tfrac{z}{2}},
\end{align*}
the (expected) conclusion that $Q_{\infty}(x)=H(x)$ for every $x\ge0$ becomes apparent.

To show that the convergence of $Q_{A}(x)$ to $H(x)$ as $A\to+\infty$ is from above, it is convenient to first rid the formula~\eqref{eq:QSD-cdf-answer-W0} for $Q_{A}(x)$ of the two Whittaker $W$ functions involved in it, and instead express it through a ratio of two modified Bessel $K$ functions (of the second kind): the Bessel functions and ratios thereof appear to have been studied more extensively than the Whittaker $W$ function. Specifically, observe that since
\begin{align}\label{eq:Whit0-BesselK-id}
W_{0,b}(2z)
&=
\sqrt{\dfrac{2z}{\pi}}\, K_{b}(z),
\end{align}
as given, e.g., by~\cite[Identity~9.6.48,~p.~377]{Abramowitz+Stegun:Handbook1964}, formula~\eqref{eq:QSD-cdf-answer-W0} can be rewritten equivalently as
\begin{align}\label{eq:QSD-cdf-answer-K}
Q_{A}(x)
&=
\begin{cases}
1,&\;\text{if $x\ge A$;}\\[2mm]
\sqrt{\dfrac{A}{x}}\,\dfrac{e^{-\tfrac{1}{x}}\,K_{\tfrac{1}{2}\xi(\lambda)}\left(\dfrac{1}{x}\right)}{e^{-\tfrac{1}{A}}\,K_{\tfrac{1}{2}\xi(\lambda)}\left(\dfrac{1}{A}\right)},&\;\text{if $x\in[0,A)$;}\\[8mm]
0,&\;\text{otherwise},
\end{cases}
\end{align}
and this formula, though completely equivalent to formula~\eqref{eq:QSD-cdf-answer-W0}, will prove to be more convenient for our purposes than~\eqref{eq:QSD-cdf-answer-W0}.

We are now ready to show that $Q_{A}(x)\ge H(x)$ for every $x\ge0$. Specifically, let us focus only on the case of $x\in(0,A)$, because the result is trivial for $x\in\{0\}\cup[A,+\infty)$. The idea is to appeal to the inequality
\begin{align}\label{eq:BesselK-ratio-upr-bnd}
\dfrac{K_{b}(x_1)}{K_{b}(x_2)}
&<
e^{x_2-x_1}\left(\dfrac{x_2}{x_1}\right)^{\tfrac{1}{2}},
\;\;
0<x_{1}<x_{2},
\;\;
-\dfrac{1}{2}<b<\dfrac{1}{2},
\end{align}
which was first stated as a conjecture by~\cite[pp.~589--590]{Baricz:PEMS2010}, but its first formal proof is apparently due to~\cite{Yang+Zheng:PAMS2017}; cf.~\cite[Inequality~(3.2),~p.~2951]{Yang+Zheng:PAMS2017}. From inequality~\eqref{eq:BesselK-ratio-upr-bnd}, formula~\eqref{eq:QSD-cdf-answer-K}, and Remark~\ref{rem:xi-complex-real} whereby $\xi(\lambda_{A})\in[0,1]$ for any $A\ge\tilde{A}\approx10.240465$, one can conclude that
\begin{align}\label{eq:QSD-cdf-lwr-bnd}
e^{\tfrac{2}{A}}\,H(x)
&<
Q_{A}(x),
\;\;
x\in(0,A),
\end{align}
which is a much tighter lowerbound for $Q_{A}(x)$ than just $H(x)$ itself, and may thus be considered a new result in its own right.
% We can now claim that
%\begin{align}\label{eq:QSD-cdf-lwr-bnd}
%\min\left\{1,e^{\tfrac{2}{A}}\,H(x)\right\}
%&\le
%Q_{A}(x),
%\;\;
%\text{for any}
%\;\;
%x\ge(0,A),
%\;\;
%\text{or}
%\;\;
%\min\left\{1,e^{\tfrac{2}{A}}\,H(x)\right\}
%&\le
%Q_{A}(x),
%\end{align}
%and the inequality is strict whenever $x\in(0,A)$; for any other $x$ the lowerbound is trivial.

Similarly, by~\cite[Inequality~(3.3),~p.~580]{Baricz:PEMS2010}, which states that
\begin{align}\label{eq:BesselK-ratio-lwr-bnd}
e^{x_2-x_1}\left(\dfrac{x_2}{x_1}\right)^{\abs{b}}
&<
\dfrac{K_{b}(x_1)}{K_{b}(x_2)},
\;\;
0<x_{1}<x_{2},
\;\;
-\dfrac{1}{2}<b<\dfrac{1}{2},
\end{align}
and formula~\eqref{eq:QSD-cdf-answer-K}, we find that
\begin{align}\label{eq:QSD-cdf-upr-bnd}
Q_{A}(x)
&\le
\min\left\{1,e^{\tfrac{2}{A}}\,H(x)\left(\dfrac{A}{x}\right)^{\tfrac{1}{2}-\tfrac{1}{2}\abs{\xi(\lambda_{A})}}\right\},
\;\;
x\ge0,
\end{align}
which, as will be demonstrated shortly, is a fairly stringent upperbound, especially if $A$ is large; this bound is a new result in itself, too.

It is evident that together inequalities~\eqref{eq:QSD-cdf-lwr-bnd} and~\eqref{eq:QSD-cdf-upr-bnd} imply that $Q_{A}(x)\downarrow H(x)$ as $A\to+\infty$ for every fixed $x\ge0$. However, it is actually possible to show more, thanks to~\cite[Theorem~2.6(iii),~p.~2948]{Yang+Zheng:PAMS2017}, whereby the function
\begin{align*}
f(b)
\coloneqq
\log\left[\dfrac{K_{b}(x_1)}{K_{b}(x_2)}\right],
\;\;
\text{with}
\;\;
0<x_1<x_2
,
\;\;
b\in\mathbb{R}
\end{align*}
is convex. As we shall now prove explicitly, this recently established property of the Bessel $K$ function enables one to conclude that
\begin{align}\label{eq:QSD-part-derivA-neg}
\dfrac{\partial}{\partial A}Q_{A}(x)
&\le
0,
\end{align}
for every $x\in(0,A)$, so that $Q_{A_{1}}(x)\ge Q_{A_{2}}(x)$ for every $x\ge0$ whenever $A_{2}\ge A_{1}\ge \tilde{A}\approx 10.240465$; recall Remark~\ref{rem:xi-complex-real} and equation~\eqref{eq:tildeA-eqn-def} which defines $\tilde{A}\approx 10.240465$.

One way to find the partial derivative of $Q_{A}(x)$ with respect to $A>0$ with $x\in(0,A)$ assumed constant is via direct differentiation of formula~\eqref{eq:QSD-cdf-answer-W0}. The necessary Whittaker $W$ function derivative identity is
\begin{align}\label{eq:WhitW-deriv-formula}
\dfrac{\partial}{\partial z}\left[e^{-\tfrac{z}{2}}W_{0,b}(z)\right]
&=
-\dfrac{1}{z}\,e^{-\tfrac{z}{2}}W_{1,b}(z),
\end{align}
which is a special case of the more general identity
\begin{align*}
\dfrac{\partial}{\partial z}\left[e^{-\tfrac{z}{2}} z^{a}W_{a,b}(z)\right]
&=
-e^{-\tfrac{z}{2}} z^{a-1}W_{a+1,b}(z)
\end{align*}
given by, e.g.,~\cite[Formula~(2.4.24),~p.~25]{Slater:Book1960}; incidentally, identity~\eqref{eq:WhitW-deriv-formula} is also a way to get formula~\eqref{eq:QSD-pdf-answer} for the quasi-stationary distribution's pdf $q_{A}(x)$ from formula~\eqref{eq:QSD-cdf-answer-W0} for $Q_{A}(x)$, and vise versa. By virtue of identities~\eqref{eq:WhitW-deriv-formula} and~\eqref{eq:Whit0-BesselK-id}, and equation~\eqref{eq:lambda-eqn}, the derivative of $Q_{A}(x)$ with respect to $A$ simplifies to
\begin{align*}
\dfrac{\partial }{\partial A}
Q_{A}(x)
&=
Q_{A}(x)\left.\left\{\dfrac{\partial}{\partial b}\log\left[\left.K_{b}\left(\dfrac{1}{A}\right)\right/K_{b}\left(\dfrac{1}{x}\right)\right]\right\}\right|_{b=\tfrac{1}{2}\xi(\lambda_{A})}\dfrac{d}{dA}\left[-\dfrac{1}{2}\xi(\lambda_{A})\right],
\end{align*}
and we note the similarity of the $\log$ of the ratio of two Bessel $K$ function to the setting of~\cite{Yang+Zheng:PAMS2017}. %$q_{A}(x)$ is the quasi-stationary distribution's pdf, given by~\eqref{eq:QSD-pdf-answer} and~\eqref{eq:lambda-eqn}. Since $q_{A}(A)=0$, we further obtain
%\begin{align*}
%\dfrac{\partial }{\partial A}
%Q_{A}(x)
%&=
%Q_{A}(x)\left.\left\{\dfrac{\partial}{\partial b}\log\left[\left.K_{b}\left(\dfrac{1}{A}\right)\right/K_{b}\left(\dfrac{1}{x}\right)\right]\right\}\right|_{b=\tfrac{1}{2}\xi(\lambda_{A})}\dfrac{d}{dA}\left[-\dfrac{1}{2}\xi(\lambda_{A})\right],
%\end{align*}
Now observe that $\xi(\lambda_{A})$ defined by~\eqref{eq:xi-def} is a monotonically increasing function $A>0$, for $\lambda_{A}>0$ is a monotonically decreasing function $A>0$. Hence, if we could show that
\begin{align*}
\left.\left\{\dfrac{\partial}{\partial b}\log\left[\left.K_{b}\left(\dfrac{1}{A}\right)\right/K_{b}\left(\dfrac{1}{x}\right)\right]\right\}\right|_{b=\tfrac{1}{2}\xi(\lambda_{A})}
&\ge
0,
\;\;
x\in(0,A),
\end{align*}
then the desired conclusion~\eqref{eq:QSD-part-derivA-neg} would follow. This is where~\cite[Theorem~2.6(iii),~p.~2948]{Yang+Zheng:PAMS2017} we mentioned earlier comes in. Specifically, since $\xi(\lambda_{A})$ is between 0 and 1 for any $A\ge\tilde{A}$, we have
\begin{align*}
\left.\left\{\dfrac{\partial}{\partial b}\log\left[\left.K_{b}\left(\dfrac{1}{A}\right)\right/K_{b}\left(\dfrac{1}{x}\right)\right]\right\}\right|_{b=0}
&\le
\left.\left\{\dfrac{\partial}{\partial b}\log\left[\left.K_{b}\left(\dfrac{1}{A}\right)\right/K_{b}\left(\dfrac{1}{x}\right)\right]\right\}\right|_{b=\tfrac{1}{2}\xi(\lambda_{A})}
,
\;\;
x\in(0,A),
\end{align*}
but
\begin{align*}
\left.\left\{\dfrac{\partial}{\partial b}\log\left[\left.K_{b}\left(\dfrac{1}{A}\right)\right/K_{b}\left(\dfrac{1}{x}\right)\right]\right\}\right|_{b=\tfrac{1}{2}\xi(\lambda_{A})}
&\le
\left.\left\{\dfrac{\partial}{\partial b}\log\left[\left.K_{b}\left(\dfrac{1}{A}\right)\right/K_{b}\left(\dfrac{1}{x}\right)\right]\right\}\right|_{b=\tfrac{1}{2}}
,
\;\;
x\in(0,A),
\end{align*}
and the next question is to evaluate each of the two derivatives: one at $b=0$ and one at $b=1/2$. The derivative at $b=0$ is zero, and the reasons it is zero are two. The first is the fact that $K_{b}(z)>0$ for all $z>0$ and $b\in\mathbb{R}$; cf., e.g.,~\cite[p.~2944]{Yang+Zheng:PAMS2017}. The second reason is the identity
\begin{align*}
\left.\left[\dfrac{\partial}{\partial b}K_{b}(z)\right]\right|_{b=0}
&=
0,
\end{align*}
as given, e.g., by~\cite[Identity~1.14.2.1,~p.~40]{Brychkov:Book2008}. For the derivative at $b=1/2$ we find
\begin{align*}
\left.\left\{\dfrac{\partial}{\partial b}\log\left[\left.K_{b}\left(\dfrac{1}{A}\right)\right/K_{b}\left(\dfrac{1}{x}\right)\right]\right\}\right|_{b=\tfrac{1}{2}}
&=
e^{\tfrac{2}{A}}\E1\left(\dfrac{2}{A}\right)-e^{\tfrac{2}{x}}\E1\left(\dfrac{2}{x}\right)
\;
(>0),
\;\;
x\in(0,A),
\end{align*}
where $\E1(z)$ denotes the exponential integral function~\eqref{eq:E1-func-def}. This result is an immediate consequence of~\cite[Identity~(13),~p.~80]{Watson:Book1922} which states that
\begin{align}\label{eq:BesselK-ind-one-half}
K_{\tfrac{1}{2}}(z)
&=
\sqrt{\dfrac{\pi}{2z}}\, e^{-z},
\end{align}
and~\cite[Identity~1.14.2.3,~p.~40]{Brychkov:Book2008} which states that
\begin{align*}
\left.\left[\dfrac{\partial}{\partial b}K_{b}(z)\right]\right|_{b=\pm\tfrac{1}{2}}
&=
\mp\sqrt{\dfrac{\pi}{2z}}\, e^{z}\Ei(-2z)
=
\pm\sqrt{\dfrac{\pi}{2z}}\, e^{z}\E1(2z),
\end{align*}
where $\Ei(z)$ denotes the exponential integral function~\eqref{eq:Ei-func-def}.

At this point we can conclude that
\begin{align}\label{eq:QSD-cdf-deriv-A-dbl-ineq}
Q_{A}(x)
\left[
e^{\tfrac{2}{A}}\E1\left(\dfrac{2}{A}\right)-e^{\tfrac{2}{x}}\E1\left(\dfrac{2}{x}\right)\right]\dfrac{d}{dA}\left[-\dfrac{1}{2}\xi(\lambda_{A})\right]
&\le
\dfrac{\partial}{\partial A}Q_{A}(x)
\le
0,
\end{align}
for every $x\in(0,A)$. This shows~\eqref{eq:QSD-part-derivA-neg}, i.e., gives the desired conclusion that $Q_{A}(x)$ is a nonincreasing function of $A$ for every fixed $x\in(0,A)$, at least for $A\ge\tilde{A}\approx 10.240465$; recall, again, that $\tilde{A}\approx 10.240465$ was introduced in Remark~\ref{rem:xi-complex-real}.
\begin{remark}
While the assumption made earlier that $A\ge\tilde{A}\approx 10.240465$ is required to ensure the index shared by the two Bessel $K$ functions involved in formula~\eqref{eq:QSD-cdf-answer-K} is purely real, so that inequalities~\eqref{eq:BesselK-ratio-upr-bnd} and~\eqref{eq:BesselK-ratio-lwr-bnd} do apply, and yield the bounds~\eqref{eq:QSD-cdf-lwr-bnd} and~\eqref{eq:QSD-cdf-upr-bnd} for $Q_{A}(x)$, these bounds actually seem to remain valid for $A<\tilde{A}\approx 10.240465$ as well. Specifically, we carried out an extensive numerical experiment to test the double inequality
\begin{align}\label{eq:QSD-cdf-upr-bnd-gnrl}
\min\left\{1,e^{\tfrac{2}{A}}\,H(x)\right\}
&\le
Q_{A}(x)
\le
\min\left\{1,e^{\tfrac{2}{A}}\,H(x)\left(\dfrac{A}{x}\right)^{\tfrac{1}{2}-\tfrac{1}{2}\abs{\Re[\xi(\lambda_{A})]}}\right\},
\;\;
x\ge0,
\end{align}
and could not find a single value of $A>0$ for which the double inequality would not hold. For $A\ge\tilde{A}\approx 10.240465$, the right half of this inequality coincides with~\eqref{eq:QSD-cdf-upr-bnd}, and for $A<\tilde{A}\approx 10.240465$, it reduces down to
\begin{align*}%\label{eq:QSD-cdf-upr-bnd-gnrl}
\min\left\{1,e^{\tfrac{2}{A}}\,H(x)\right\}
&\le
Q_{A}(x)
\le
\min\left\{1,e^{\tfrac{2}{A}}\,H(x)\left(\dfrac{A}{x}\right)^{\tfrac{1}{2}}\right\},
\;\;
x\ge0,
\end{align*}
which we determined through the numerical experiment to be somewhat conservative.
\end{remark}

We implemented the obtained lower- and upper-bounds~\eqref{eq:QSD-cdf-lwr-bnd} and~\eqref{eq:QSD-cdf-upr-bnd} in a {\it Mathematica} script, and used the script to produce Figures~\ref{fig:QST-cdf-H-bnds-errs-A10},~\ref{fig:QST-cdf-H-bnds-errs-A50}, and~\ref{fig:QST-cdf-H-bnds-errs-A100}. These figures show the performance of the bounds as functions of $x\in[0,A]$ for $A=10$, $50$, and $100$; for $A=10$, which is less than $\tilde{A}\approx 10.240465$, the upperbound was computed using~\eqref{eq:QSD-cdf-upr-bnd-gnrl}. Specifically, Figures~\ref{fig:QST-cdf-H-bnds-A10},~\ref{fig:QST-cdf-H-bnds-A50}, and~\ref{fig:QST-cdf-H-bnds-A100} show the quasi-stationary distribution's cdf $Q_{A}(x)$, the stationary cdf $H(x)$, and the lower- and upper-bounds~\eqref{eq:QSD-cdf-lwr-bnd} and~\eqref{eq:QSD-cdf-upr-bnd}---all as functions of $x\in[0,A]$, for $A=10$, $50$, and $100$, respectively. To solve equation~\eqref{eq:lambda-eqn} and compute $\lambda_{A}$, and, subsequently, also evaluate $Q_{A}(x)$, we relied on the {\it Mathematica} script written earlier by~\cite{Polunchenko:SA2017a}. The figures leave the impression that the bounds perform reasonably well, even though in quickest change-point detection any value of $A$ that is less than $100$ is generally considered low because the corresponding false alarm risk level in high. The only ``exception'' is the case of $A=10$: the bounds in this case are somewhat loose, especially the upperbound, whose corresponding relative deviation from $Q_{A}(x)$ peaks almost $100\,\%$. However, expectedly, as $A$ increases, the quality of the bounds improves, and the gap between $Q_{A}(x)$ and $H(x)$ diminishes as well, uniformly for all $x\in[0,A]$. For $A=50$ and higher, either bound's largest margin of error is adequate. Juxtaposed side-by-side to Figures~\ref{fig:QST-cdf-H-bnds-A10},~\ref{fig:QST-cdf-H-bnds-A50}, and~\ref{fig:QST-cdf-H-bnds-A100} are Figures~\ref{fig:QST-cdf-bnds-err-A10},~\ref{fig:QST-cdf-bnds-err-A50}, and~\ref{fig:QST-cdf-bnds-err-A100}  which show the corresponding lower- and upper-bound errors. Specifically, the lowerbound error is defined as the excess of $Q_{A}(x)$ over the bound, while the upperbound error is defined as the excess of the bound over $Q_{A}(x)$. All three error figures suggest that the upperbound is generally tighter than the lowerbound, unless $A$ is very low, in which case both bounds are off by a large margin.
\begin{figure}[h!]
    \centering
    \begin{subfigure}{0.48\textwidth}
        \centering
        \includegraphics[width=\linewidth]{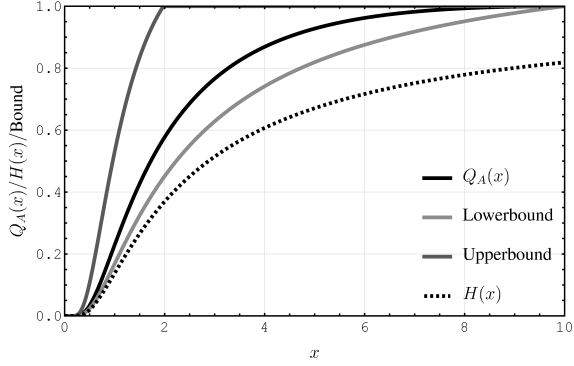}
        \caption{$Q_{A}(x)$, the lower- and upper-bounds, and $H(x)$.}
        \label{fig:QST-cdf-H-bnds-A10}
    \end{subfigure}
    \hspace*{\fill}
    \begin{subfigure}{0.48\textwidth}
        \centering
        \includegraphics[width=\linewidth]{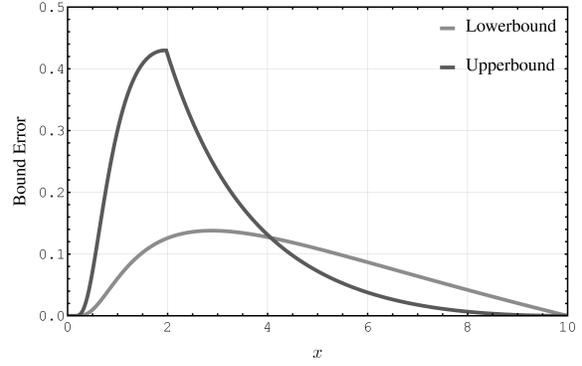}
        \caption{Corresponding lower- and upper-bound errors.}
        \label{fig:QST-cdf-bnds-err-A10}
    \end{subfigure}
    \caption{Quasi-stationary distribution's cdf, $Q_{A}(x)$, its lower- and upper-bounds, their corresponding errors, and stationary distribution's cdf, $H(x)$---all as functions of $x\in[0,A]$ for $A=10$.}
    \label{fig:QST-cdf-H-bnds-errs-A10}
\end{figure}
\begin{figure}[h!]
    \begin{subfigure}{0.48\textwidth}
        \centering
        \includegraphics[width=\linewidth]{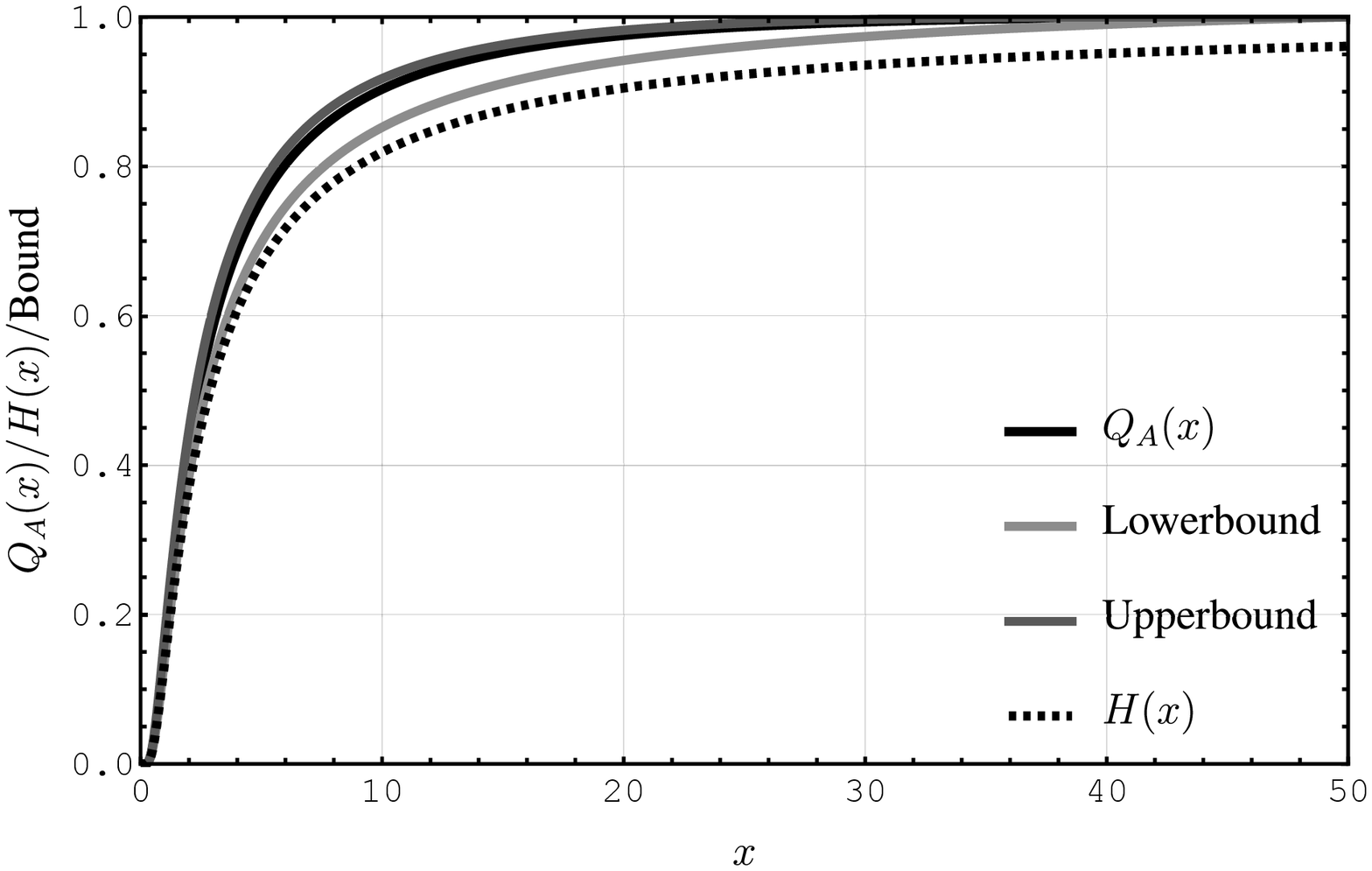}
        \caption{$Q_{A}(x)$, the lower- and upper-bounds, and $H(x)$.}
        \label{fig:QST-cdf-H-bnds-A50}
    \end{subfigure}
    \hspace*{\fill}
    \begin{subfigure}{0.48\textwidth}
        \centering
        \includegraphics[width=\linewidth]{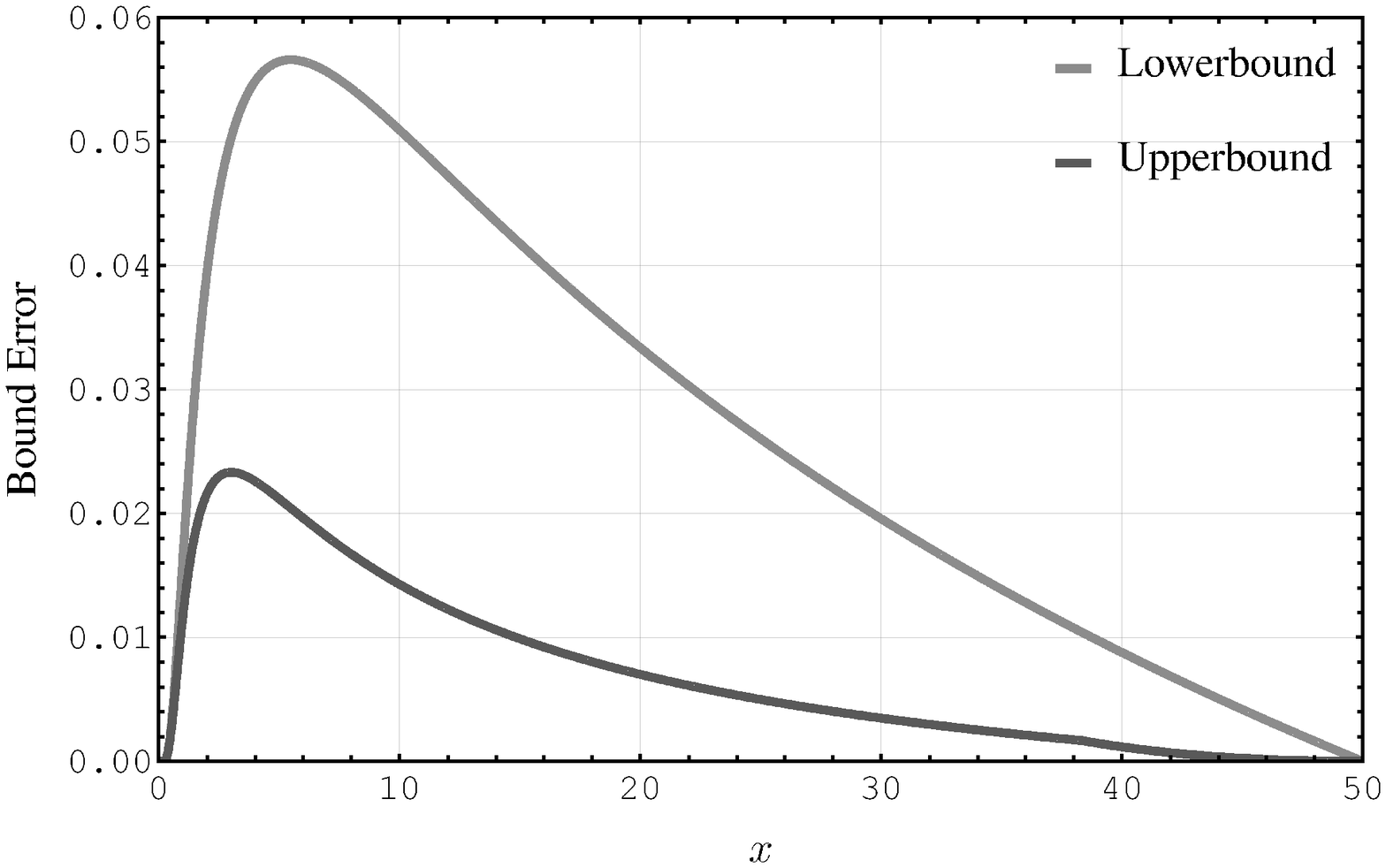}
        \caption{Corresponding lower- and upper-bound errors.}
        \label{fig:QST-cdf-bnds-err-A50}
    \end{subfigure}
    \caption{Quasi-stationary distribution's cdf, $Q_{A}(x)$, its lower- and upper-bounds, their corresponding errors, and stationary distribution's cdf, $H(x)$---all as functions of $x\in[0,A]$ for $A=50$.}
    \label{fig:QST-cdf-H-bnds-errs-A50}
\end{figure}
\begin{figure}[h!]
    \begin{subfigure}{0.48\textwidth}
        \centering
        \includegraphics[width=\linewidth]{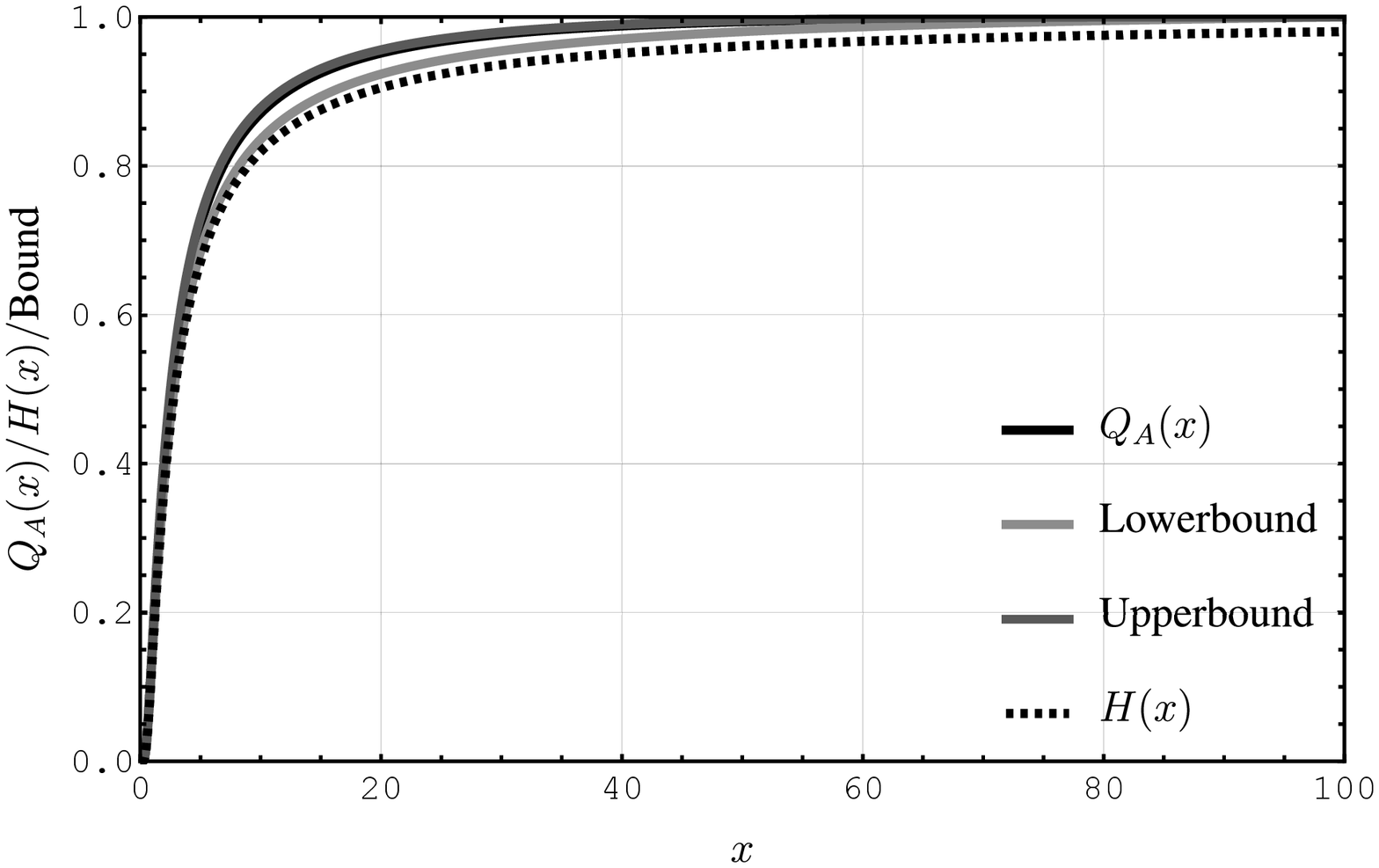}
        \caption{$Q_{A}(x)$, the lower- and upper-bounds, and $H(x)$.}
        \label{fig:QST-cdf-H-bnds-A100}
    \end{subfigure}
    \hspace*{\fill}
    \begin{subfigure}{0.48\textwidth}
        \centering
        \includegraphics[width=\linewidth]{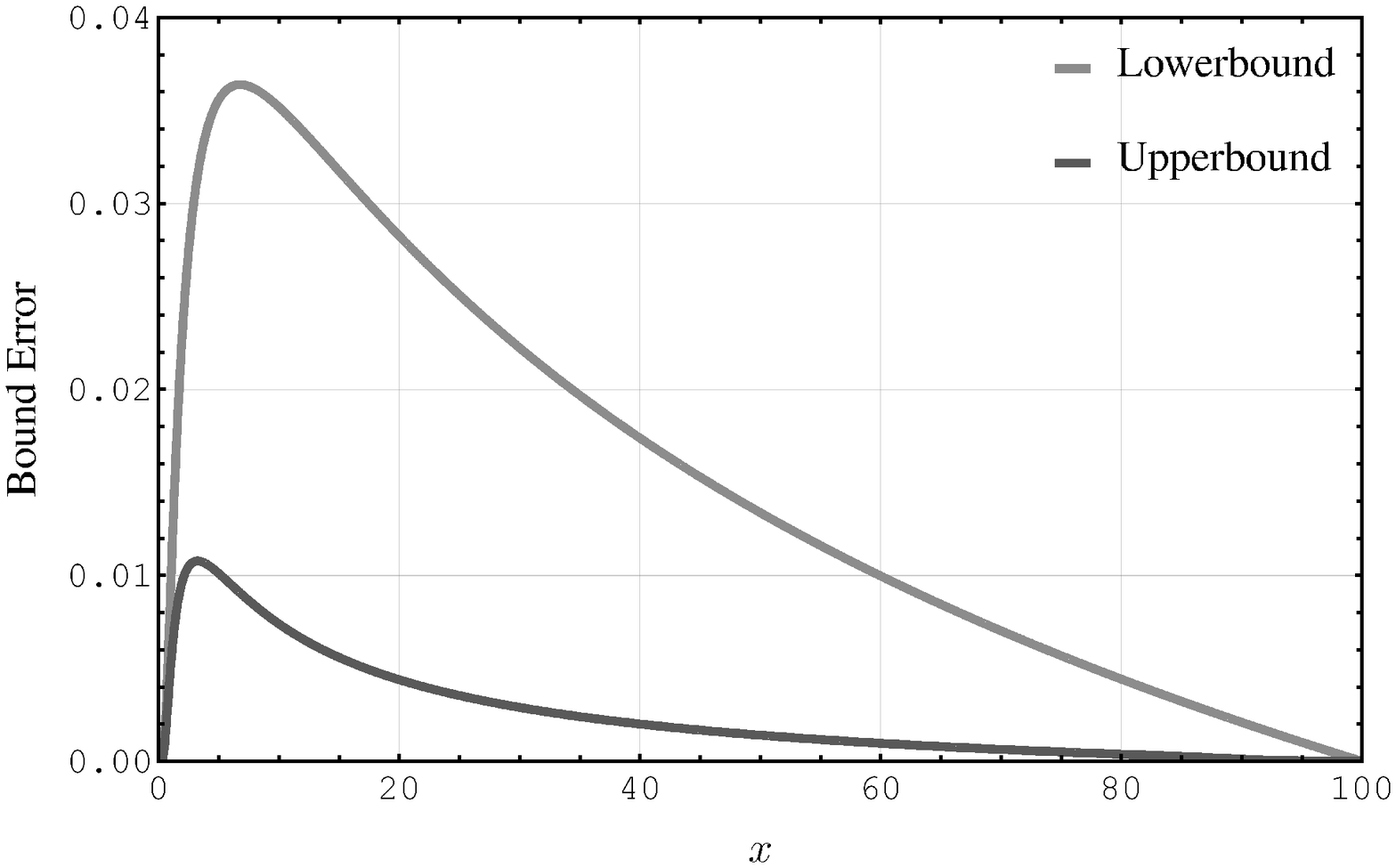}
        \caption{Corresponding lower- and upper-bound errors.}
        \label{fig:QST-cdf-bnds-err-A100}
    \end{subfigure}
    \caption{Quasi-stationary distribution's cdf, $Q_{A}(x)$, its lower- and upper-bounds, their corresponding errors, and stationary distribution's cdf, $H(x)$---all as functions of $x\in[0,A]$ for $A=100$.}
    \label{fig:QST-cdf-H-bnds-errs-A100}
\end{figure}

We are now in position to explain the main contribution of this work. It concerns the actual speed at which $Q_{A}(x)$ descends down to $H(x)$, as $A\to+\infty$, for every $x\ge0$, and can be formally and succinctly stated thus:
\begin{align*}
\sup_{x\ge0}\left[Q_{A}(x)-H(x)\right]
&=
O\left(\dfrac{\log(A)}{A}\right)
,
\;\;
\text{as $A\to+\infty$},
\end{align*}
and it is shown next.

Our proof of the foregoing asymptotics relies on the inequality
\begin{align*}
Q_{A}(x)-H(x)
&\le
H(x)\left\{\left(e^{\tfrac{2}{A}}-1\right)+e^{\tfrac{2}{A}}\left(\dfrac{A}{x}\right)^{\tfrac{1}{2}-\tfrac{1}{2}\xi(\lambda_{A})}\left[e^{\tfrac{2}{A}}\E1\left(\dfrac{2}{A}\right)-e^{\tfrac{2}{x}}\E1\left(\dfrac{2}{x}\right)\right]\dfrac{1-\xi(\lambda_{A})}{2}\right\},
\end{align*}
valid for any $x\in[0,A]$ and at least for $A\ge\tilde{A}\approx10.240465$; recall Remark~\ref{rem:xi-complex-real} and that $\tilde{A}$ is the solution of equation~\eqref{eq:tildeA-eqn-def}. From this inequality and the lowerbound~\eqref{eq:QSD-cdf-lwr-bnd} it can be seen that $Q_{A}(x)$ descends down to $H(x)$ no faster than $O(1/A)$ but no slower than $O(\log(A)/A)$. The upper estimate for the convergence speed is because
\begin{align*}
e^{\tfrac{2}{A}}
-
1
&=
O\left(\dfrac{1}{A}\right)
,
\;\;
\text{as $A\to+\infty$},
\end{align*}
which is trivial. The lower estimate is because
\begin{align*}
e^{\tfrac{2}{A}}\left(\dfrac{A}{x}\right)^{\tfrac{1}{2}-\tfrac{1}{2}\xi(\lambda_{A})}\left[e^{\tfrac{2}{A}}\E1\left(\dfrac{2}{A}\right)-e^{\tfrac{2}{x}}\E1\left(\dfrac{2}{x}\right)\right]\dfrac{1-\xi(\lambda_{A})}{2}
&=
O\left(\dfrac{\log(A)}{A}\right),
\;\;
\text{as $A\to+\infty$},
\end{align*}
and to see this observe that $1-\xi(\lambda_{A})=O(1/A)$, as can be concluded from~\eqref{eq:lambda-dbl-ineq} plugged into~\eqref{eq:xi-def}, and recall that $\lim_{x\to+\infty}\big(\sqrt[x]{x}\,\big)=1$ and that
\begin{align*}
\dfrac{1}{2}\log\left(1+\dfrac{2}{z}\right)
&\le
e^{z}\E1(z)
\le
\log\left(1+\dfrac{1}{z}\right),
\end{align*}
as given by~\cite[Inequality~5.1.20,~p.~229]{Abramowitz+Stegun:Handbook1964};

To prove the above inequality for $Q_{A}(x)-H(x)$, fix $A\ge\tilde{A}\approx10.240465$ and $x\in(0,A)$, and Taylor-expand $Q_{A}(x)$ given by~\eqref{eq:QSD-cdf-answer-K} regarded as a function of $\xi(\lambda)/2\in[0,1/2)$ near $1/2$ up to the linear term; the assumption made here that $\xi(\lambda_{A})$ is nonnegative is not restrictive, because $Q_{A}(x)$ is an even function of $\xi(\lambda_{A})$, as can be inferred from formula~\eqref{eq:QSD-cdf-answer-W0} and Remark~\ref{rem:xi-symmetry}. On account of~\eqref{eq:BesselK-ratio-lwr-bnd} and its consequence~\eqref{eq:BesselK-ind-one-half} this gives
\begin{align*}
\begin{split}
Q_{A}(x)
&=
e^{\tfrac{2}{A}}H(x)+\left.\sqrt{\dfrac{A}{x}}\dfrac{e^{-\tfrac{1}{x}}}{e^{-\tfrac{1}{A}}}\left\{\left[\left.K_{b}\left(\dfrac{1}{x}\right)\right/K_{b}\left(\dfrac{1}{A}\right)\right]\dfrac{\partial}{\partial b}\log\left[\left.K_{b}\left(\dfrac{1}{x}\right)\right/K_{b}\left(\dfrac{1}{A}\right)\right]\right\}\right|_{b=b^{*}}\left(\dfrac{1}{2}\xi(\lambda_{A})-\dfrac{1}{2}\right)
\\
&=
e^{\tfrac{2}{A}}H(x)+
\left.\sqrt{\dfrac{A}{x}}\dfrac{e^{-\tfrac{1}{x}}}{e^{-\tfrac{1}{A}}}\left\{\left[\left.K_{b}\left(\dfrac{1}{x}\right)\right/K_{b}\left(\dfrac{1}{A}\right)\right]\dfrac{\partial}{\partial b}\log\left[\left.K_{b}\left(\dfrac{1}{A}\right)\right/K_{b}\left(\dfrac{1}{x}\right)\right]\right\}\right|_{b=b^{*}}\dfrac{1-\xi(\lambda_{A})}{2},
\end{split}
\end{align*}
for some $b^{*}=b^{*}(x,A)\in(\xi(\lambda)/2,1/2)$. Now, since the second term in the obtained Taylor expansion is nonneative, the expansion can be easily turned into a new upperbound for $Q_{A}(x)$: it suffices to apply~\eqref{eq:QSD-cdf-upr-bnd} along with~\cite[Theorem~2.6(iii),~p.~2948]{Yang+Zheng:PAMS2017} to upperbound the expression dependent on $b^{*}$ on the right of the Taylor expansion. Specifically, the new upperbound for $Q_{A}(x)$ takes the form
\begin{align*}
Q_{A}(x)
&\le
\min\left\{1,e^{\tfrac{2}{A}}H(x)\left\{1+\left(\dfrac{A}{x}\right)^{\tfrac{1}{2}-\tfrac{1}{2}\xi(\lambda_{A})}\left[e^{\tfrac{2}{A}}\E1\left(\dfrac{2}{A}\right)-e^{\tfrac{2}{x}}\E1\left(\dfrac{2}{x}\right)\right]\dfrac{1-\xi(\lambda_{A})}{2}\right\}\right\},
\;
x\in(0,A),
\end{align*}
and it is generally tighter than the upperbound~\eqref{eq:QSD-cdf-upr-bnd} we obtained earlier.

To conclude, we remark that one can improve the upperbound for $Q_{A}(x)-H(x)$ by Taylor-expanding $Q_{A}(x)$ given by~\eqref{eq:QSD-cdf-answer-K} regarded as a function of $\xi(\lambda)/2\in[0,1/2)$ near $1/2$ further, viz. up to the quadratic term. Specifically, a simple calculation using~\eqref{eq:QSD-cdf-upr-bnd} and~\cite[Theorem~2.6(iii),~p.~2948]{Yang+Zheng:PAMS2017} gives
\begin{align*}
\begin{split}
Q_{A}(x)
&\le
e^{\tfrac{2}{A}}H(x)\,\Biggl\{1+\left[e^{\tfrac{2}{A}}\E1\left(\dfrac{2}{A}\right)-e^{\tfrac{2}{x}}\E1\left(\dfrac{2}{x}\right)\right]\dfrac{1-\xi(\lambda_{A})}{2}
+
\\
&\qquad\qquad\qquad
+
\dfrac{1}{2}\left(\dfrac{A}{x}\right)^{\tfrac{1}{2}-\tfrac{1}{2}\xi(\lambda_{A})}\left[e^{\tfrac{2}{A}}\E1\left(\dfrac{2}{A}\right)-e^{\tfrac{2}{x}}\E1\left(\dfrac{2}{x}\right)\right]^2\left(\dfrac{1-\xi(\lambda_{A})}{2}\right)^2
\Biggr\},
\;\;
x\in(0,A),
\end{split}
\end{align*}
where again $A\ge\tilde{A}\approx10.240465$ with $\tilde{A}\approx10.240465$ defined by equation~\eqref{eq:tildeA-eqn-def} and Remark~\ref{rem:xi-complex-real}.

%+-----------------------------------------------------------------------------------------------+%
\section{Conclusion and outlook}
\label{sec:conclusion}

The obtained new bounds for the cdf of the quasi-stationary distribution of the Generalized Shiryaev--Roberts process do have applications in quickest change-point detection. For one, the upperbounds can be used to tightly cap Pollak's~\citeyearpar{Pollak:AS85} maximal Average Detection Delay (ADD) delivered by Pollak's~\citeyearpar{Pollak:AS85} randomized Shiryaev--Roberts--Pollak (SRP) procedure, and subsequently show that the SRP procedure is nearly minimax in the sense of Pollak~\citeyearpar{Pollak:AS85}; the rate of convergence to the unknown optimal minimax performance can be estimated as well. See, e.g.,~\cite{Polunchenko:TPA2017} who upperbounded the quasi-stationary cdf by unity, which is, of course, a very conservative upperbound. More importantly, the upperbounds and the lowerbound offered in this work may also be instrumental in showing that the Generalized Shiryaev--Roberts procedure with a carefully designed headstart is, too, nearly minimax in the sense of Pollak~\citeyearpar{Pollak:AS85}. We are of the view that any proof of such a strong claim would certainly merit a separate paper. For example,~\cite{Tartakovsky+etal:TPA2012} confirm the claim to be valid in the discrete-time setting. In the continuous-time setting, the main challenge is the fact that while all of the relevant performance characteristics can be computed analytically and in a closed form, the expressions are quite involved, and depend on special functions; see, e.g.,~\cite{Polunchenko:SA2016}. However, the bounds obtained in this work may enable one to find an explicit, simpler yet sufficiently tight upperbound for Pollak's~\citeyearpar{Pollak:AS85} maximal ADD delivered by the GSR procedure, and then use the upperbound to show that its excess over the unknown optimal minimax ADD is negligible in the limit, as the Average Run Length to false alarm goes to infinity. Some preparatory work in this direction is already underway.

%+-----------------------------------------------------------------------------------------------+%
%+-----------------------------------------------------------------------------------------------+%
%+-----------------------------------------------------------------------------------------------+%

%+-----------------------------------------------------------------------------------------------+%
\section*{Acknowledgement}
The effort of A.S.~Polunchenko was partially supported by the Simons Foundation via a Collaboration Grant in Mathematics under Award \#\,304574.

% Reference list
%+-----------------------------------------------------------------------------------------------+%
%\bibliographystyle{sqa}
%\bibliography{main,physics,special-functions,finance,stochastic-processes,differential-equations}

\end{document}